\documentclass[%
 reprint,
 amsmath,amssymb,
 aps,
]{revtex4-2}

\usepackage[utf8]{inputenc}
\usepackage{color}
\usepackage{graphicx}
\usepackage{dcolumn}
\usepackage{bm}

\begin{document}

\preprint{APS/123-QED}

\title{Alignment-Controlled Optical Orbital Trapping of Single Airborne Aerosols for Dynamical Particle Sensing}


\author{Chun-Yen Wen}
\author{Yang-Yi Lee}
\author{Chung-Lin Chao}
\author{Ruei-Ying Jian}
\author{Tzu-Ling Chen}
 \email{tlc@nycu.edu.tw}
\affiliation{%
Department of Photonics, College of Electrical and Computer Engineering, National Yang Ming Chiao Tung University, Hsinchu 30010, Taiwan\\
}%

\author{Wayne Cheng-Wei Huang}
\affiliation{
Department of Physics, National Tsing Hua University, Hsinchu 30013, Taiwan\\
}%


\date{\today}

\begin{abstract}
Optical forces in focused-beam traps are generally nonconservative, yet the controlled use of this nonconservative component for airborne single-particle dynamics remains limited. 
We demonstrate a dual-beam optical trap in which a single aerosol can be switched between localized confinement and sustained orbital motion by tuning the relative positions of two counter-propagating foci. 
The axial separation $\Delta D$ controls the onset of nonconservative circulation, while the lateral offset $\Delta R$ tunes the projected orbit size and causes a monotonic change in the rotation frequency.
T-matrix optical force calculations and Langevin simulations support this interpretation by showing that finite axial misalignment activates a circulating force component, whereas near-zero axial separation gives a confinement-dominated force field. 
Experiments confirm the predicted switching behavior through mean-square displacement and frequency measurements. 
We further show that the projected orbit geometry provides a particle-dependent observable, with the orbit anisotropy $A_y/A_x$ varying systematically with aerosol diameter. 
The results provide a compact, low-power platform for controlled orbital dynamics of single airborne particles and for future aerosol measurements based on nonequilibrium trajectory observables.
\end{abstract}

\maketitle


\section{Introduction}

Optical tweezers are widely used to confine and manipulate microscopic objects, enabling precision force measurements, microrheology, and single-particle spectroscopy~\cite{gieseler2021optical,gao2017optical,gong2018optical,bustamante2021optical}. In many applications, the trapped object is treated as a Brownian probe localized near a potential minimum, so that information is extracted from position fluctuations, trap stiffness, or the response to external perturbations~\cite{gieseler2021optical,pesce2015step,volpe2013simulation}. However, optical forces are not generally conservative: radiation pressure and scattering can generate circulating components of the force field, allowing sustained circulation or deterministic orbital motion rather than only static confinement~\cite{nan2022creating,zhou2025longitudinal}.

Orbital optical trapping has been realized using structured beams, optical vortices, time-dependent beam scanning, and misaligned counter-propagating beams~\cite{yang2021optical,chen2016characteristics,chen2016dynamics,li2018dynamic,raj2022orbital}. Recent air-based demonstrations have further shown rapid orbital motion by aligning optical momentum transfer with the particle trajectory~\cite{droby2025optical}. Most of these studies emphasize large orbital trajectories, efficient momentum transfer, or particle sizes for which ray-optics or largely deterministic force models capture the dominant dynamics. Here we instead focus on a compact-orbit regime more relevant to single-aerosol measurements: micron-sized airborne particles close to the confinement-to-rotation boundary. This regime is advantageous because modest alignment changes can reversibly switch the particle between stable confinement and orbital motion, while the particle-size range of interest overlaps directly with that most relevant to atmospheric aerosol processes~\cite{seinfeld2016atmospheric,poschl2005atmospheric,krieger2012exploring}.

Counter-propagating dual-beam traps are particularly suitable here because opposing radiation pressures provide axial confinement without a substrate or liquid medium, while allowing flexible control of aerosol size and orbital state.
In our system, the relative beam geometry is controlled by two independent parameters: the axial separation $\Delta D$ and the lateral offset $\Delta R$ between the two foci. Finite $\Delta D$ breaks the longitudinal symmetry of the counter-propagating beams and produces a circulating nonconservative component of the optical force field, while $\Delta R$ tunes the accessible orbit size. The aerosols studied here lie in an intermediate Mie-scattering regime, where particle size, projected orbital amplitude, focal offsets, and Brownian fluctuations of the trajectory are all comparable on micrometer length scales. As a result, the confinement-to-rotation transition is especially sensitive to the local topology of the nonconservative force field.

\begin{figure*}[t]
  \centering
  \includegraphics[width=0.9\linewidth]{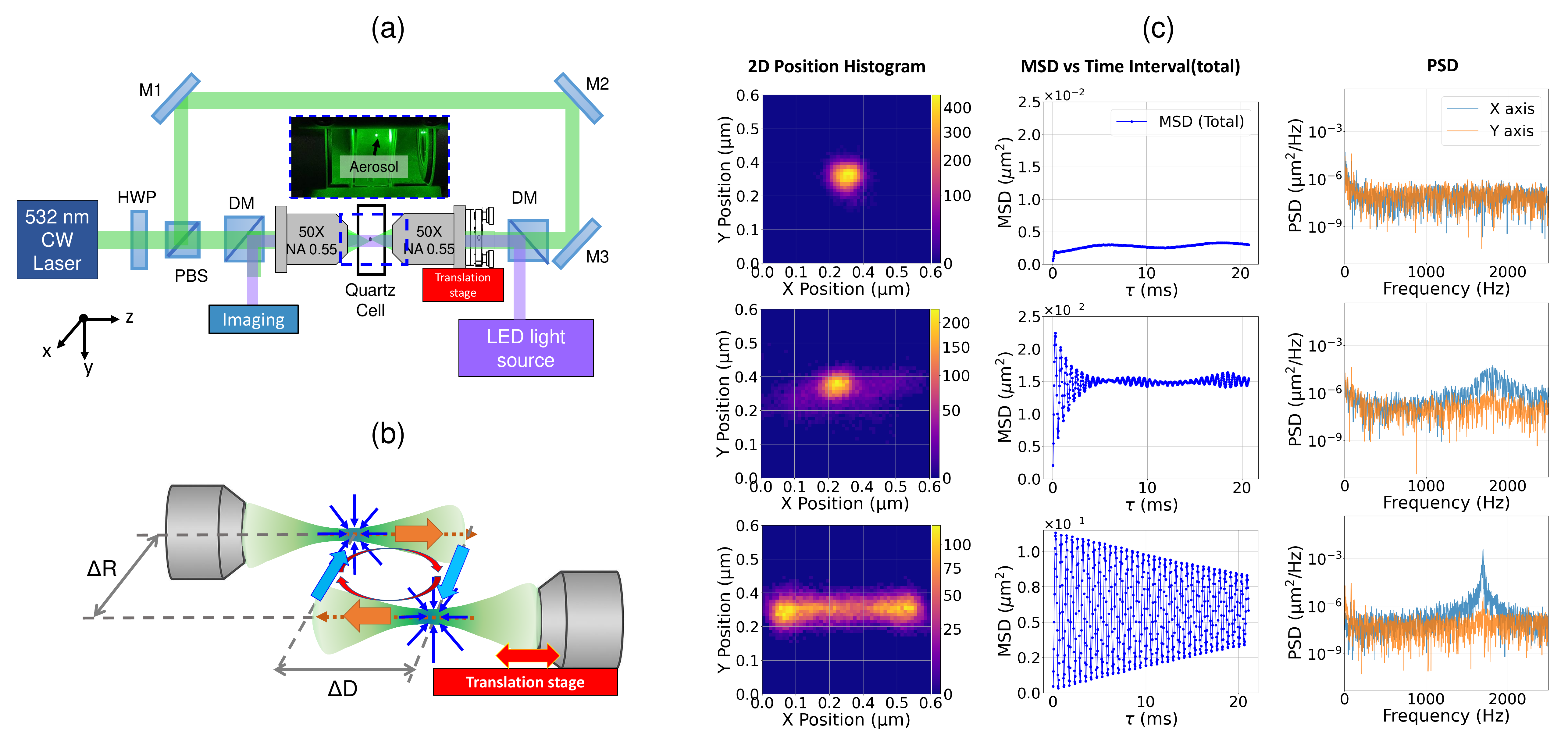}
\caption{
\textbf{Dual-beam optical trap, alignment control, and experimental signatures of orbital motion.}
(a) Experimental setup of the free-space counter-propagating dual-beam trap for single airborne aerosols. A 532~nm CW laser is split into two balanced trapping arms and focused into the center of a quartz cell by two opposing $50\times$ objectives. Bright-field imaging is provided by LED illumination coupled coaxially into the trapping region.
(b) Schematic of the relative focal geometry. The lateral offset $\Delta R$ and axial separation $\Delta D$ between the two foci define the two alignment parameters that control the optical force landscape. Finite $\Delta D$ produces a scattering-force imbalance, while $\Delta R$ modifies the projected orbit size.
(c) Representative experimental signatures of three motional states in the imaging plane: tight confinement (top), small-orbit rotation (middle), and large-orbit rotation (bottom). Left: two-dimensional position histograms. Center and Right: corresponding total MSD as a function of lag time $\tau$ and and PSD. The confined state shows a compact distribution and a plateauing MSD, whereas rotating states exhibit elongated projected trajectories and oscillatory MSD traces with increasing modulation amplitude for larger orbits.
}
  \label{fig:setup}
\end{figure*}

Here we combine high-speed position tracking, optical force-field calculations, and Langevin simulations to map the confinement-to-rotation transition in the $(\Delta R,\Delta D)$ alignment space. We show that axial scattering and transverse gradient forces cooperate to generate compact nonconservative orbits, and that the orbit geometry provides a size-sensitive dynamical observable for airborne aerosols. In particular, the gravity-axis orbit projection decreases systematically with particle diameter, providing a complementary readout when apparent particle size from direct imaging is affected by projection or defocus. These results establish compact orbital trapping as a controllable nonequilibrium state and a promising route toward trajectory-based single-aerosol sensing.

\section{Dual-beam optical orbital trap for single aerosols}
\label{sec:experiment}

\subsection{Experimental setup and alignment control}
\label{subsec:setup}

The experimental platform is shown in Fig.~\ref{fig:setup}(a). 
A continuous-wave 532~nm laser is divided into two counter-propagating arms and focused into the center of a quartz trapping cell by two identical $50\times$ microscope objectives (NA\,0.55). 
The optical power in the two arms is adjusted with a half-wave plate and a polarizing beam splitter and is balanced before the trapping objectives. 
The single-arm power before the trapping objective is typically 105~mW, corresponding to approximately 75--85~mW delivered after the objective and trapping cell.
The measured beam waist in the trapping region is approximately $0.7~\mu$m, as determined by a knife-edge measurement described in Appendix~\ref{app:exp_calibration}.

The relative positions of the two foci define the two experimental control parameters. 
As illustrated in Fig.~\ref{fig:setup}(b), $\Delta R$ is the lateral offset between the foci, and $\Delta D$ is their axial separation along the beam-propagation direction. 
These offsets determine the local optical force field experienced by the aerosol. 
The lateral offset $\Delta R$ modifies the transverse force balance and the accessible orbit size, whereas the axial separation $\Delta D$ changes the longitudinal overlap of the focal regions and the balance of the counter-propagating scattering forces.

Aerosols are introduced into the cell using a nebulizer-based delivery line. 
After a single particle is trapped, the flow is restricted to reduce gas-flow perturbations during trajectory measurements. 
Particle motion is recorded by coaxial bright-field imaging: blue LED illumination is coupled into the trapping region through a dichroic mirror, and residual 532~nm trapping light is rejected by a notch filter before detection. 
High-speed camera images are used to reconstruct the two-dimensional particle trajectory. 
From these trajectories we compute position histograms, MSD, power spectra, projected orbit amplitudes, and rotation frequencies.

\subsection{Experimental signatures of confinement and orbital motion}
\label{subsec:state_signature}

Representative motional states observed in the same dual-beam geometry are shown in Fig.~\ref{fig:setup}(c). 
When the two foci are nearly balanced, the aerosol remains localized near the confinement center, where the MSD rapidly approaches a plateau, as expected for confined Brownian motion.
With increasing focal misalignment, the position distribution broadens and develops an elongated structure. At the same time, the MSD acquires oscillatory features, indicating that the particle periodically revisits positions along a closed trajectory. 
For larger orbital states, the projected histogram expands further and the MSD oscillation amplitude increases substantially. 

These observations show that the dual-beam system can access a continuous set of motional states, ranging from tight confinement to sustained orbital motion, as the focal alignment is varied. The MSD traces and PSD provide a simple experimental signature that distinguishes localized confinement, small-orbit rotation, and large-orbit rotation. The next question is whether this transition reflects only a gradual broadening of the trap or the onset of a qualitatively different, nonconservative force-field topology. To address this, we compare the measurements with calculated optical force fields and Langevin simulations.

\section{Optical-force calculation and nonconservative origin of orbital motion}
\label{sec:model}

\begin{figure*}[t]
\centering
\includegraphics[width=\linewidth]{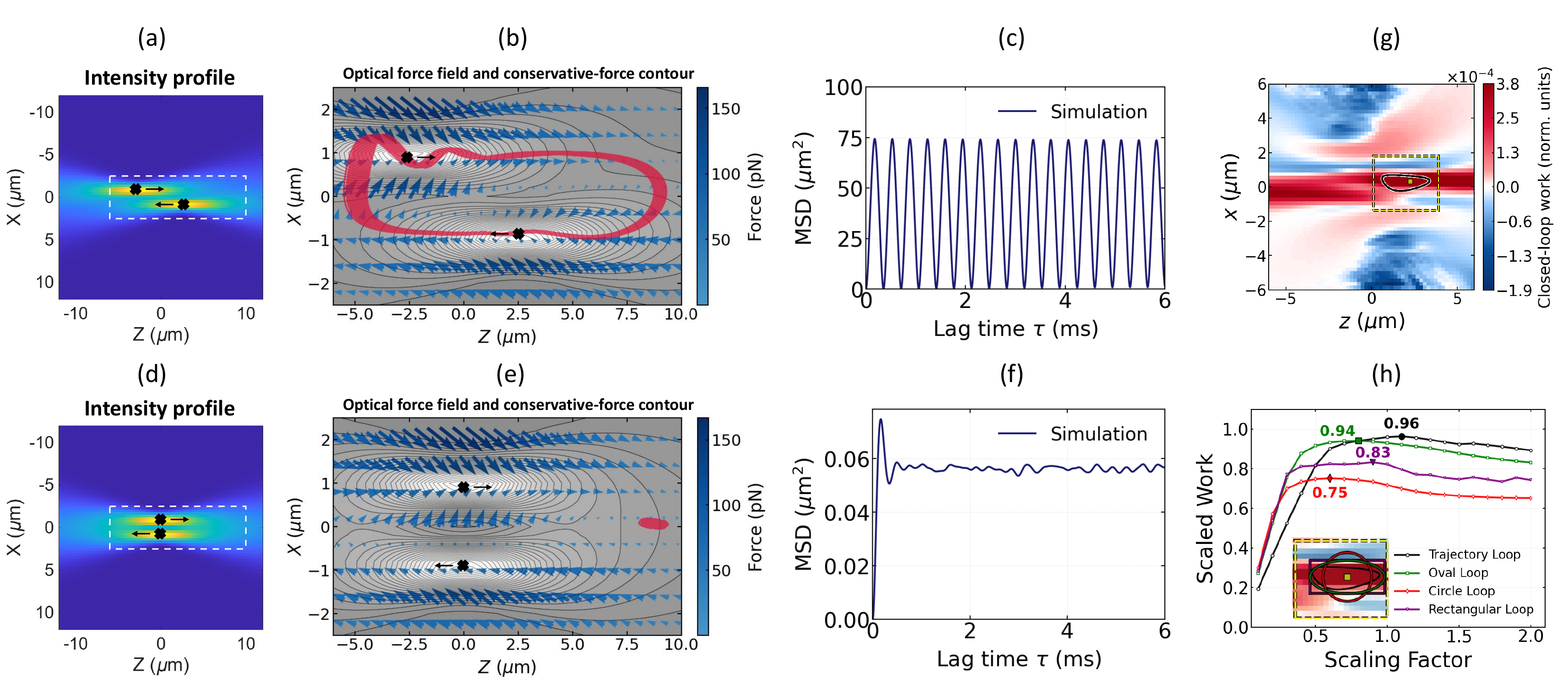}
\caption{
\textbf{Nonconservative optical force field underlying aerosol orbital motion.}
(a--c) Representative rotating condition with finite axial offset.
(a) Intensity profile of the two counter-propagating beams in the analysis plane.
(b) Calculated total optical force field (arrows, colored by force magnitude) overlaid with contours of the local restoring landscape; the magenta curve shows a representative simulated trajectory.
(c) Corresponding simulated MSD, which exhibits oscillatory behavior characteristic of orbital motion.
(d--f) Representative confinement condition with near-zero axial offset. The force field is dominated by local confinement, and the simulated MSD rapidly approaches a plateau.
(g) Closed-loop work map obtained by translating a fixed elliptical contour across the calculated force field. Positive work near the simulated orbit indicates net optical work over one cycle.
(h) Scaled closed-loop work for different contour shapes. The trajectory-like loop gives the largest value, showing that the simulated orbit follows the strongest circulating part of the nonconservative force field.
}
\label{fig:mechanism}
\end{figure*}

To interpret this confinement-to-rotation transition, we calculate the optical force field using a T-matrix optical tweezers model.
The trapped aerosols have diameters in the micrometer range and therefore lie in an intermediate Mie regime at $\lambda=532~\mathrm{nm}$, where neither Rayleigh-limit dipole forces nor simple geometric-optics models are sufficient. The T-matrix calculation gives the total optical force
$\mathbf{F}_{\mathrm{opt}}(\mathbf{r})$
for the two focused, counter-propagating Gaussian beams at prescribed values of $\Delta R$ and $\Delta D$.

The calculated force field is used both to visualize the local optical-force landscape and as the deterministic input for particle-trajectory simulations. Because the total optical force is generally nonconservative, the effective confinement contours shown below are used only to visualize the local restoring landscape and should not be interpreted as a global scalar potential. The particle dynamics are modeled with an inertial Langevin equation,
\begin{equation}
m\ddot{\mathbf{r}}+\gamma \dot{\mathbf{r}}
=
\mathbf{F}_{\mathrm{opt}}(\mathbf{r})
+
\boldsymbol{\xi}(t),
\end{equation}
where $m$ is the particle mass, $\gamma=6\pi\eta a$ is the Stokes drag coefficient for a particle of radius $a$ in air of viscosity $\eta$, and $\boldsymbol{\xi}(t)$ is a thermal noise force satisfying
\begin{equation}
\langle \xi_i(t)\xi_j(t')\rangle
=
2k_B T\gamma\,\delta_{ij}\delta(t-t').
\end{equation}
The force field is tabulated on a Cartesian grid and interpolated during trajectory integration. For each alignment condition, simulated trajectories are analyzed in the same manner as the experimental data, including the trajectory, MSD, and PSD. The rotation frequency is extracted from the dominant spectral peak of the particle trajectory.
Including inertia lowers the simulated rotation frequency relative to the overdamped approximation and improves the agreement with experiment.
The simulations are used primarily to identify the force-field mechanism and to determine how changes in $\Delta D$ and $\Delta R$ modify the transition between confinement and orbital motion. Additional details of the optical parameters, force normalization, numerical integration, and reduced-dimensional treatment are given in Appendix~\ref{app:simulation}.

Figure~\ref{fig:mechanism} compares two representative alignment conditions with finite and near-zero axial offset. For finite axial separation [Fig.~\ref{fig:mechanism}(a--c)], the displaced focal planes break the longitudinal symmetry of the counter-propagating beams. The resulting force field contains a circulating component in the analysis plane: transverse restoring forces keep the particle near the beam-overlap region, while the imbalance of the counter-propagating scattering forces drives motion around a closed path. Langevin simulations using this force field produce a stable orbit and an oscillatory MSD, consistent with the experimental signatures of rotational motion described above.

To quantify this circulating component, we evaluate the closed-loop work
\[
W_{\mathrm{loop}}
=
\oint_{\mathcal{C}}
\mathbf{F}_{\mathrm{opt}}(\mathbf{r})\cdot d\mathbf{r},
\]
where $\mathcal{C}$ is a closed contour in the calculated force field (with more details in Appendix~\ref{app:close}). For a purely conservative force field, this integral vanishes for any closed contour. A positive value of $W_{\mathrm{loop}}$ indicates that the optical force performs net work over one cycle and can therefore sustain dissipative orbital motion. As shown in Fig.~\ref{fig:mechanism}(g), the closed-loop work is positive in the rotational region and is maximized near the simulated trajectory. To examine the dependence on contour choice, we repeated the calculation for several trial loop shapes over a range of scaling factors. Figure~\ref{fig:mechanism}(h) shows that the trajectory-like contour gives the largest scaled work, whereas simpler circular or rectangular loops yield smaller values. This supports our interpretation that the observed orbit is sustained by the strongest nonconservative circulation in the calculated force landscape, rather than serving as a stand-alone predictor of the realized trajectory.

By contrast, when the axial separation is nearly zero [Fig.~\ref{fig:mechanism}(d--f)], the scattering forces from the two beams remain approximately balanced in the trapping region. In this case, the force field is dominated by local confinement, the simulated trajectory remains localized, and the MSD rapidly approaches a plateau without sustained oscillation. The corresponding closed-loop work is much reduced (not shown here), showing that the circulating nonconservative component is no longer sufficient to support a stable orbit.

These results identify axial focal separation as the key control parameter that activates nonconservative circulation in the dual-beam trap. Stable orbital motion is not simply a geometric broadening of the trap, but a dynamical state sustained by scattering-driven work and closed by transverse restoring confinement. Related dual-beam force-field simulations have likewise emphasized the role of scattering-force imbalance in generating closed particle trajectories~\cite{zhou2025longitudinal}.

\section{Alignment-space control of aerosol orbital trapping}
\label{sec:alignment_control}

\begin{figure*}[t]
\centering
\includegraphics[width=\linewidth]{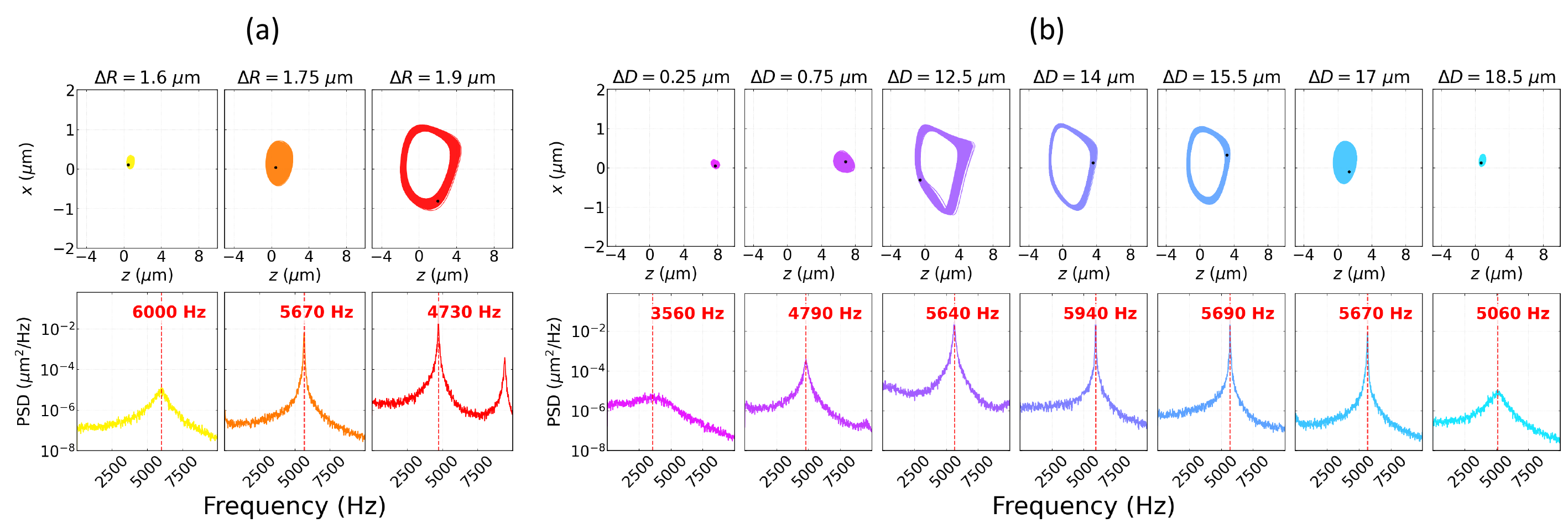}
\caption{
\textbf{Simulated trajectory evolution under alignment control.}
(a) Simulated trajectories and corresponding power spectral densities (PSDs) at fixed $\Delta D=17~\mu$m while varying $\Delta R$. Increasing $\Delta R$ enlarges the orbit mainly along the transverse direction and reduces the rotation frequency.
(b) Simulated trajectories and corresponding PSDs at fixed $\Delta R=1.75~\mu$m while varying $\Delta D$. Increasing $\Delta D$ first activates and strengthens the circulation, then gradually broadens the orbit and lowers the frequency. For $\Delta D \gtrsim 18~\mu$m, the rotation peak broadens and the orbit becomes less well defined, indicating weaker orbital stabilization.
}
\label{fig:sim_alignment}
\end{figure*}

Having identified axial focal separation as the origin of the nonconservative circulation, we next examine how the orbital state is controlled by the axial separation $\Delta D$ and the lateral offset $\Delta R$. With the laser power fixed, tuning $(\Delta R,\Delta D)$ allows the same dual-beam platform to operate either as a stable confining trap or as an orbital trap. In the rotating state, the camera records the projected particle trajectory in the $x$--$y$ imaging plane. We characterize each orbit by the projected amplitudes $A_x$ and $A_y$, while the rotation frequency is extracted from the dominant spectral peak of the position trace. The projected amplitudes describe the orbit geometry, whereas the spectral peak provides a reproducible dynamical measure for comparing different alignment conditions.

Before turning to the experimental scans, Fig.~\ref{fig:sim_alignment} summarizes the corresponding trends in the simulated trajectories. At fixed $\Delta D=17~\mu$m [Fig.~\ref{fig:sim_alignment}(a)], increasing $\Delta R$ enlarges the orbit mainly along the transverse direction, so that the projected amplitude in $x$ increases while the rotation frequency decreases. At fixed $\Delta R=1.75~\mu$m [Fig.~\ref{fig:sim_alignment}(b)], increasing $\Delta D$ first activates and strengthens the orbital motion, producing faster circulation, and then broadens the orbit while gradually reducing the rotation frequency. For larger axial separations (e.g., $\Delta D\gtrsim 18~\mu$m), the spectral peak broadens and the orbit becomes less well defined, indicating that the circulating state is no longer sharply stabilized. These simulated trends provide a compact dynamical picture of the two alignment controls: $\Delta R$ mainly sets the orbit size, whereas $\Delta D$ governs the onset and strength of the circulating drive.

\begin{figure}[t]
\centering
\includegraphics[width=0.85\linewidth]{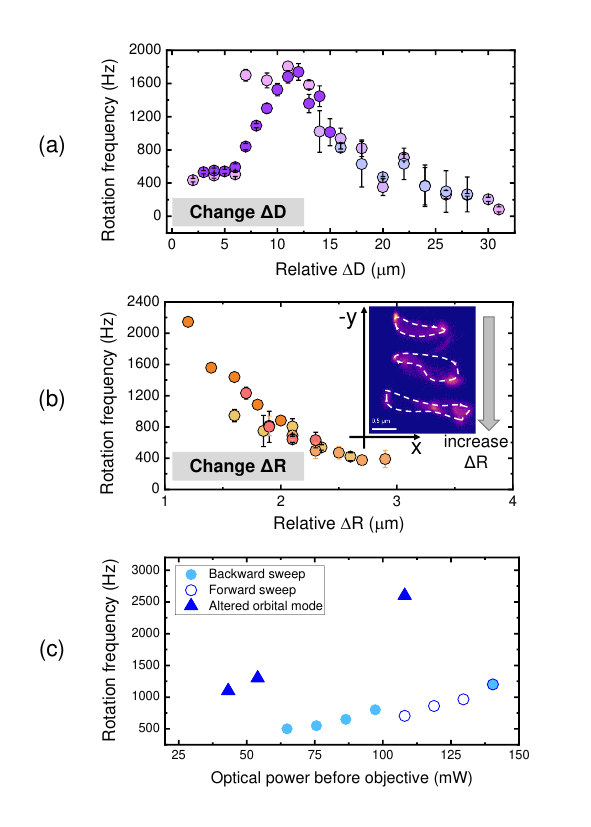}
\caption{
\textbf{Experimental alignment and power control of aerosol orbital motion.}
(a) Measured rotation frequency as the axial focal separation $\Delta D$ is varied at nearly fixed $\Delta R$. Increasing $\Delta D$ first activates and strengthens the orbital motion, whereas larger axial separation broadens the orbit and reduces the frequency.
(b) Measured rotation frequency as the lateral focal offset $\Delta R$ is varied at nearly fixed $\Delta D$. Increasing $\Delta R$ enlarges the accessible orbit (inset) and decreases the rotation frequency.
(c) Rotation frequency as a function of trapping power for forward and backward power sweeps. The monotonic increase of frequency with power indicates that stronger optical forcing increases the circulating drive while maintaining the orbital state.
}
\label{fig:alignment_control}
\end{figure}

The experimental tuning behavior is summarized in Fig.~\ref{fig:alignment_control}. At nearly fixed $\Delta R$, increasing $\Delta D$ first drives the particle from localized confinement into a stable rotating state [Fig.~\ref{fig:alignment_control}(a)]. This behavior is consistent with the force-field mechanism in Fig.~\ref{fig:mechanism}, where the axial separation breaks the longitudinal symmetry of the counter-propagating beams and produces a finite circulating component in the optical force field. As $\Delta D$ is increased further, the orbit frequency decreases, which indicates that the circulating drive does not grow monotonically with axial separation; instead, the particle moves through a more extended region where the effective restoring landscape is weaker. At sufficiently large $\Delta D$, the orbit becomes less sharply defined, consistent with the broadened spectral peak seen in the simulations [Fig.~\ref{fig:sim_alignment}].

At nearly fixed $\Delta D$, varying $\Delta R$ primarily changes the projected orbit size [Fig.~\ref{fig:alignment_control}(b)]. Increasing $\Delta R$ shifts the two beam axes farther apart and enlarges the accessible trajectory. The measured rotation frequency decreases with increasing $\Delta R$, consistent with a longer path length and reduced spatial overlap between transverse confinement and scattering-driven circulation. In agreement with the simulations, the main geometric effect of increasing $\Delta R$ is an increase in the transverse orbit amplitude.

We further examine the role of optical power at a fixed orbital configuration. As shown in Fig.~\ref{fig:alignment_control}(c), the rotation frequency increases monotonically with trapping power for both forward and backward sweeps. This behavior is consistent with the optical force scaling approximately with incident power [Eq.~(B1) in Appendix~\ref{app:simulation}], so that higher power strengthens the circulating drive and increases the orbital speed. Under the same alignment condition, we also observe frequency shifts for a distinct altered orbital mode, but these data follow the same overall trend.

These measurements establish a practical control rule for the dual-beam orbital trap. Finite $\Delta D$ is required to activate the circulating nonconservative force, whereas $\Delta R$ mainly selects the accessible orbit size and therefore the rotation frequency. Optical power then provides a secondary control over the circulation rate once a stable orbit is established. The orbital state is therefore not an accidental consequence of misalignment, but a controllable state in alignment space. These controls also provide the basis for using the orbit as a readout: once a stable trajectory is formed, the trapped particle is characterized not only by its mean position, but also by its orbit size, orbit anisotropy, and rotation frequency.

\section{Orbital dynamics as a sensing channel}
\label{sec:sensing}

We next ask whether the controlled orbital state contains information about the trapped particle. 
A rotating aerosol provides two classes of observables. 
The projected orbit geometry reflects the force balance sampled by the particle over one cycle, whereas the rotation frequency reports the instantaneous rate of circulation. 
As shown below, these quantities are affected differently by particle size and by transient changes in the trapping condition.

\subsection{Size-dependent orbital geometry}
\label{sec:size_geometry}

To examine size-dependent changes in the orbital trajectory, we measured rotating aerosols over a range of diameters. In each case, the particle was first trapped in a nonrotating confined state at finite $\Delta D$ and then driven into a regular rotating state by tuning the alignment through $\Delta R$. Even under nominally similar alignment conditions (that is, comparable relative $\Delta R$), the absolute rotation frequency alone is not a sufficiently robust metric for comparing different particles. Although the frequency shows some correlation with particle size, the trend is not monotonic across alignment groups. We therefore focus instead on the projected orbit amplitudes $A_x$ and $A_y$, extracted from the reconstructed trajectory in the imaging plane.

\begin{figure*}[t]
\centering
\includegraphics[width=\linewidth]{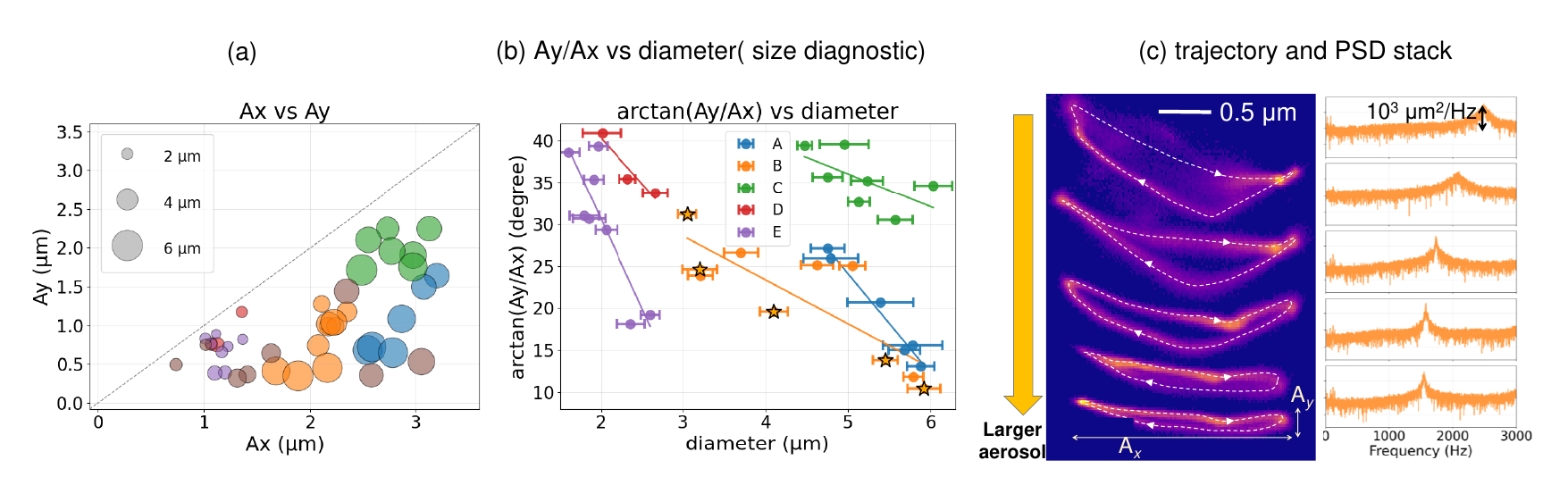}
\caption{
\textbf{Projected orbit geometry as a size-sensitive dynamical observable.}
(a) Projected orbit amplitudes $A_x$ and $A_y$ extracted from experimentally realized rotating trajectories. Marker size denotes particle diameter, and color denotes the alignment group. The distribution of points shows that the absolute projected orbit size depends on both the trapping condition and the particle size, while the rotating states remain anisotropic in the imaging plane.
(b) Orbit anisotropy expressed as $\theta=\arctan(A_y/A_x)$ as a function of particle diameter. Within each alignment group, $\theta$ decreases systematically with increasing diameter, indicating that larger particles exhibit a reduced orbit projection along the gravity axis relative to the transverse direction. Stars mark the representative trajectories displayed in panel (c).
(c) Representative reconstructed trajectories used to define $A_x$ and $A_y$, and the corresponding PSD. The dashed outlines indicate the projected orbit envelope, and the white arrows show the direction of motion along the orbit. From top to bottom, the particle diameter increases, illustrating the progressive flattening of the projected orbit along the gravity direction.
}
\label{fig:sensing}
\end{figure*}

Figure~\ref{fig:sensing}(a) shows the measured correlation between $A_x$ and $A_y$ for aerosols spanning a range of particle diameters. 
The accessible orbit size depends on the alignment condition, as expected from the results in Fig.~\ref{fig:alignment_control}. 
A more systematic trend appears when the orbit anisotropy is considered. 
We define
\[
\theta=\arctan\!\left(\frac{A_y}{A_x}\right).
\]
As shown in Fig.~\ref{fig:sensing}(b), $\theta$ decreases with increasing particle diameter within multiple experimental groups. 
Thus larger particles tend to have flatter projected orbits, with a smaller gravity-axis component relative to the transverse component. 
Representative stacked trajectories in Fig.~\ref{fig:sensing}(c) show the same behavior directly: the vertical amplitude decreases, while the horizontal extent changes less strongly within a given group.

This trend is consistent with a gravitational-loading picture. 
A larger and heavier particle samples a shifted operating point in the optical force landscape. 
The local restoring force along the gravity direction is then modified, reducing the accessible vertical orbit amplitude. 
The proportionality is not expected to be universal, because it depends on the local alignment and on the optical force field sampled by the orbit. 
Nevertheless, the repeated decrease of $\theta$ with particle diameter indicates that the projected orbit geometry carries particle-dependent information.

The geometry-based observable is useful because direct bright-field sizing of a levitated aerosol can be affected by projection, defocus, and depth-of-field effects. 
The orbit anisotropy, expressed through $A_y/A_x$ or $\theta$, provides an independent dynamical observable that can be compared across repeated alignment groups. 
In the present work this observable is used as a particle-dependent indicator rather than as an absolute size calibration.

\begin{figure*}[t]
\centering
\includegraphics[width=0.7\linewidth]{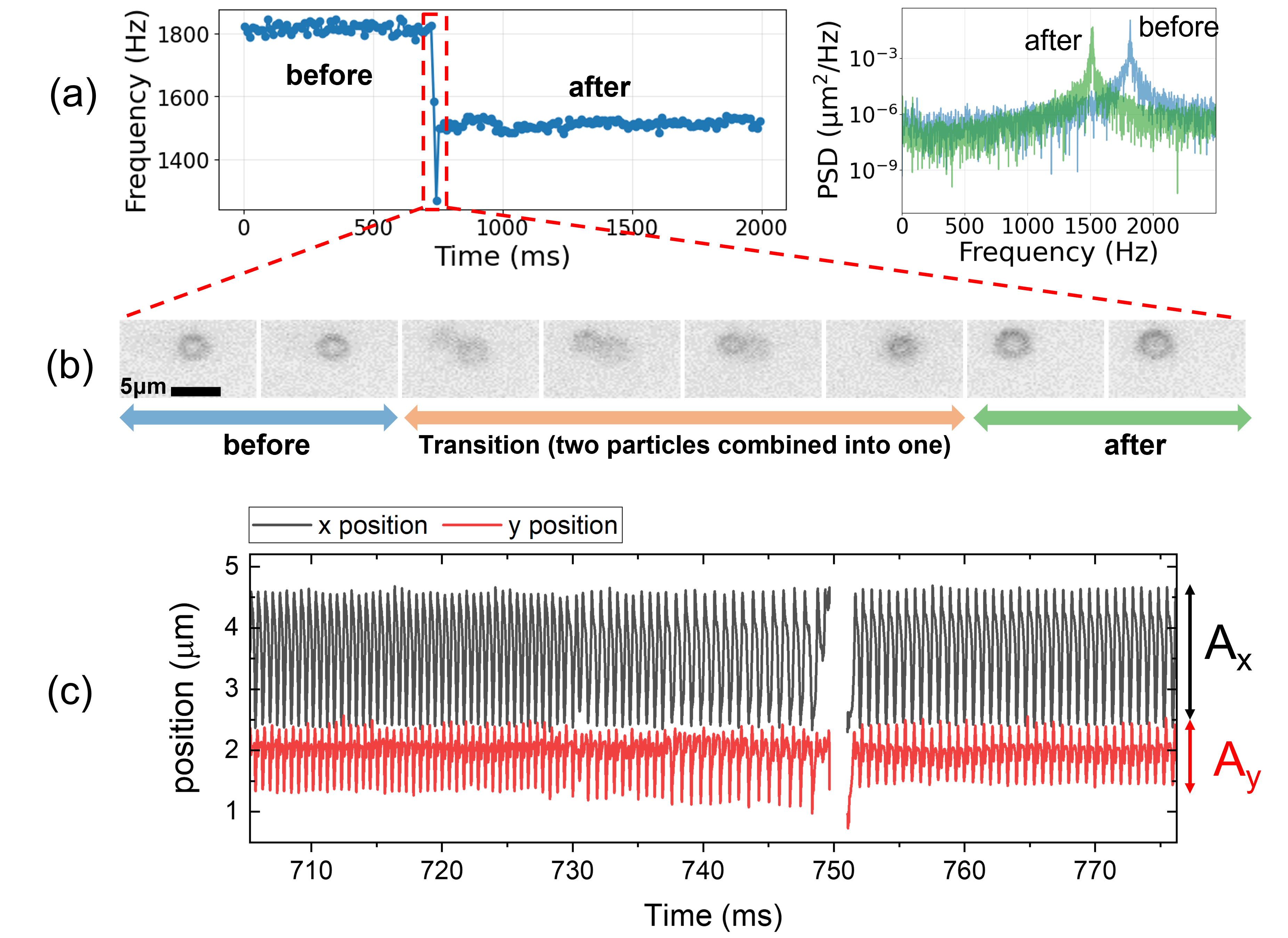}
\caption{
\textbf{Time-resolved orbital response during a particle-merging event.}
(a) Rotation frequency extracted from the dominant spectral peak of the particle trajectory as a function of time. A sudden transition occurs near $t\approx 750$~ms, where the orbital frequency shifts from the initial state to a lower-frequency state.
Right: Power spectra before and after the transition, showing the corresponding shift of the dominant rotation peak.
(b) Representative particle images recorded during the event. The enlarged image sequence shows that two nearby particles merge into a single larger particle during the transition.
(c) Simultaneously recorded $x$ and $y$ position traces in the time window marked in (a). After the merger, the projected orbit changes anisotropically: the gravity-axis amplitude $A_y$ changes clearly, whereas the transverse amplitude $A_x$ remains nearly unchanged. Together, these data show that the orbital state provides a time-resolved dynamical readout of abrupt changes in particle size and mass.
}
\label{fig:frequency_response}
\end{figure*}
\subsection{Time-resolved frequency response}

The orbital frequency provides a complementary observable for tracking time-dependent changes of a single trapped aerosol. Figure~\ref{fig:frequency_response} shows a representative transient event in which the dominant rotation frequency changes abruptly during the measurement. The extracted peak shifts from $1818.85~\mathrm{Hz}$ before the transition to $1519.52~\mathrm{Hz}$ afterward, and the corresponding PSDs are shown in Fig.~\ref{fig:frequency_response}(a).

The image sequence in Fig.~\ref{fig:frequency_response}(b) indicates that this transition coincides with a particle-merging event: two nearby particles combine into a single larger particle. The simultaneously recorded position traces [Fig.~\ref{fig:frequency_response}(c)] show that the orbital motion changes immediately after the merger. In particular, the projected orbit changes anisotropically: the gravity-axis amplitude $A_y$ is modified clearly, whereas the transverse amplitude $A_x$ remains nearly unchanged. This behavior is consistent with the interpretation developed above, namely that the gravity-axis component of the orbit is more sensitive to particle size or loading than the transverse component.

This event illustrates the distinction between the two orbit-based readouts. The projected orbit geometry is useful for comparing size-dependent behavior across repeated measurements, while the rotation frequency is well suited to following time-dependent changes of a single orbit. Although the present measurement does not yet constitute a calibrated aerosol sensor, it demonstrates that abrupt changes in particle state can be tracked in real time through simultaneous changes in frequency and orbit geometry.

\section{Discussion}

A useful feature of the present regime is that the particle size, orbital amplitude, and focal offsets are all comparable on micrometer length scales. As a result, the particle does not simply move through a fixed large-scale force landscape; instead, different parts of the orbit sample appreciably different local force-field structures, especially in the nonconservative component. In this sense, the relevant quantity is not only the force magnitude at a single point, but the topology of the force field sampled over the full trajectory. This also helps explain why the dynamics are more difficult to simulate quantitatively in the compact-orbit aerosol regime than in larger-orbit systems~\cite{droby2025optical}, where the force landscape varies more gradually and simpler deterministic or inertial descriptions can often capture the main behavior. In the present case, continuous changes in particle position modify the locally sampled force-field topology and therefore the realized orbit, which is why we use the term ``topology'' and why full trajectory calculations remain important.

Rather than emphasizing quantitative prediction of the rotation frequency or pursuing the highest attainable frequencies, as in some previous studies, the present work focuses on an alignment-controlled optical orbital trap for single airborne particles and on how this controllable orbital state can be used for aerosol sensing. The same low-power dual-beam geometry supports both stable confinement and orbital motion, with the transition controlled by two experimentally accessible parameters. In particular, $\Delta D$ activates or suppresses the nonconservative circulating drive, whereas $\Delta R$ mainly sets the accessible orbit size and therefore the rotation frequency. A simple scaling picture,
\begin{equation}
  f \sim \frac{F_\theta}{\gamma\,L_{\mathrm{orb}}},
\end{equation}
captures the observed trends: $\Delta D$ changes the effective azimuthal drive $F_\theta$, while $\Delta R$ changes the orbital path length $L_{\mathrm{orb}}$. Although this is not a complete model of the limit cycle, it explains why larger projected orbits generally rotate more slowly.

Beyond state control, the orbital state provides observables not available in a static trap. In the present measurements, the projected orbit geometry, especially $A_y/A_x$ or $\theta=\arctan(A_y/A_x)$, is more reproducible as a size-dependent quantity than the rotation frequency alone. This is expected because the frequency depends simultaneously on optical drive, drag, orbit size, and inertial effects, whereas the gravity-axis component of the orbit is more directly influenced by the vertical force balance sampled by the particle.

Several limitations and extensions should be noted. The size-dependent trends reported here are obtained from different trapping events and alignment groups, so a calibrated sensing protocol will require tracking the same particle as its size or composition changes and comparing the orbital observables with an independent measurement. In addition, the simulations assume a spherical particle with an effective refractive index. Variations in refractive-index contrast, weak absorption, or particle asphericity can shift the boundaries between confinement and rotation by modifying both the restoring and circulating components of the optical force. Finally, although the inertial Langevin description improves the agreement between simulated and measured rotation frequencies at atmospheric pressure, reduced-pressure operation will not necessarily simplify the interpretation. Lower damping may reduce Brownian broadening and increase the orbital signal-to-noise ratio, but it will also allow the nonconservative optical work to accumulate more efficiently, so that inertia and limit-cycle stability become increasingly important in setting both the orbit amplitude and the escape boundary. In that sense, the present stability diagram should be regarded as specific to a finite-damping airborne regime, rather than as a direct guide to the near-vacuum limit. Future experiments combining controlled humidity, evaporation, or photothermal excitation with independent size or composition readout will therefore be needed to turn these orbit observables into calibrated aerosol measurements.

\section{Conclusions}

We have demonstrated an alignment-controlled optical orbital trap for single airborne aerosols in a free-space dual-beam geometry. By independently tuning the lateral and axial focal offsets, $(\Delta R,\Delta D)$, and combining experiment with T-matrix force calculations and inertial Langevin simulations, we identified the force-field origin of the confinement-to-rotation transition and established how the orbital state evolves across alignment space. Finite axial separation activates a scattering-driven nonconservative circulation, while transverse restoring forces close the trajectory into a stable orbit. The resulting orbital motion is therefore a controllable dynamical state of the trap rather than an incidental consequence of misalignment.

Beyond state control, we show that the orbital motion provides trajectory-based observables that are sensitive to particle properties. In particular, the projected orbit geometry in the imaging plane varies systematically with particle size, with the gravity-axis component providing a more reproducible size-dependent signal than the rotation frequency alone. This geometry-based readout is especially useful for airborne particles, where direct image-based sizing can be distorted by projection, defocus, and depth-of-field limitations. Therefore, these results establish dual-beam aerosol orbital trapping as both a controllable nonequilibrium optical state and a promising dynamical readout for future single-particle aerosol measurements.

\begin{acknowledgments}
This work was supported by grants NSTC113-2628-M-A49-002, NSTC114-2628-M-A49-001 as well as by the Ministry of Education in Taiwan under the Yushan Young Scholar Program.
W.C.H. gratefully acknowledges funding supports from NSTC 113-2112-M-007-022-MY3 and NSTC 114-2811-M-007-047.

\end{acknowledgments}

Data availability: processed traces and derived metrics will be provided as CSV;

Code availability: Python scripts for force-map interpolation, Langevin simulation, MSD/PSD analysis, and work-map computation (with version of OTT and key parameters).

\appendix

\section{Experimental calibration and trajectory analysis}
\label{app:exp_calibration}

\begin{figure}[t]
  \centering
  \includegraphics[width=0.78\linewidth]{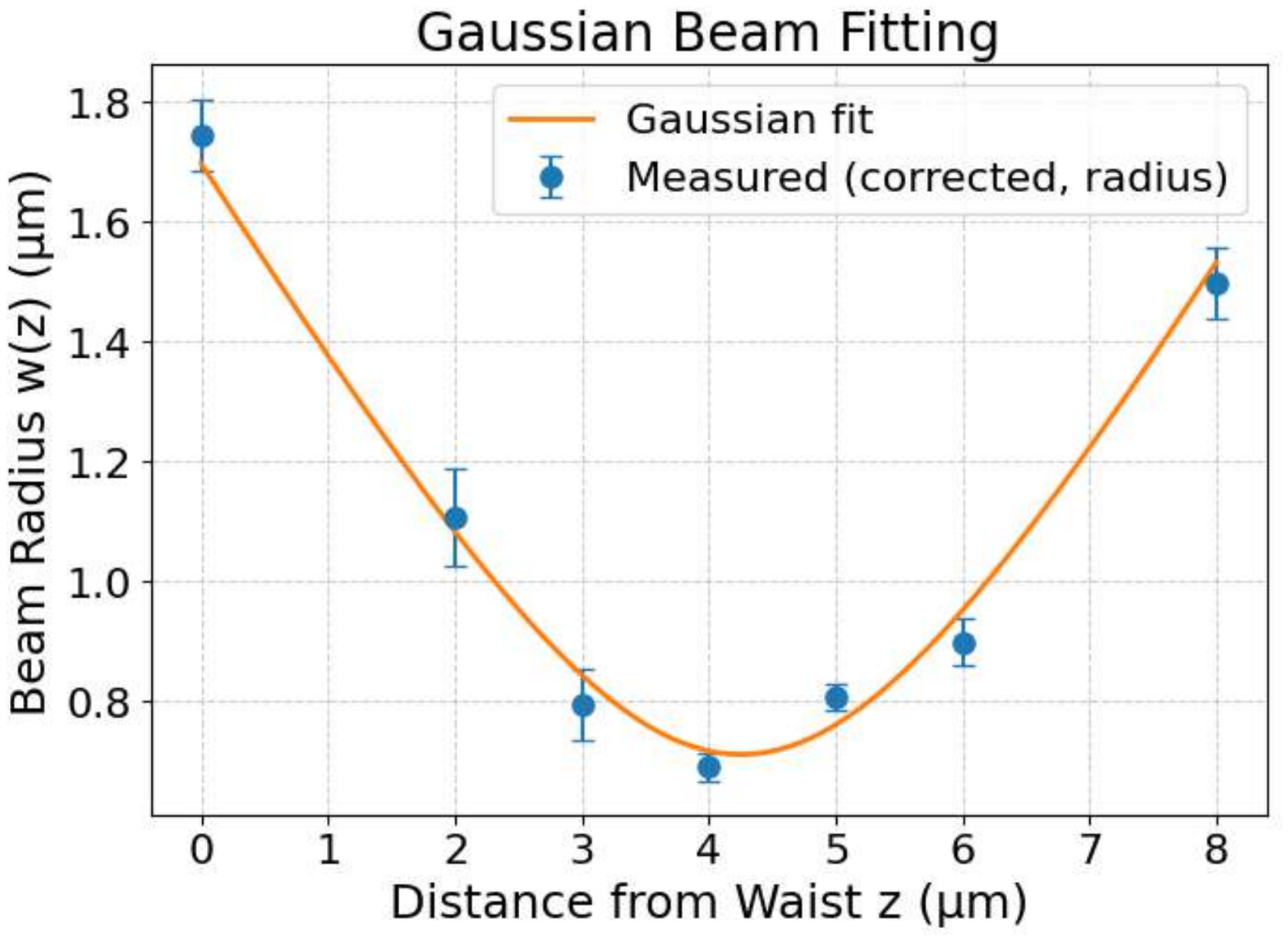}
  \caption{
  \textbf{Knife-edge beam characterization.}
  Measured beam radius \(w(z)\) obtained from knife-edge scans. 
  The edge response was fitted with the integrated Gaussian profile to extract the \(1/e^2\) radius.
  }
  \label{fig:knife}
\end{figure}

The beam waist in the trapping region was measured by a knife-edge scan. 
The transmitted power was recorded as a sharp edge was translated across the focused beam, and the response was fitted with the integrated Gaussian profile. 
The measured waist was approximately $0.7~\mu$m. 
Figure~\ref{fig:knife} shows the measured beam radius obtained from the knife-edge scans.

The possible chief-ray tilt introduced by lateral translation of the trapping objective was estimated as
\begin{equation}
\theta \lesssim \frac{\Delta R}{f},
\end{equation}
where $f\approx4~\mathrm{mm}$ is the focal length of the objective. 
For the experimental range of $\Delta R$, this gives $\theta \leq 0.04^\circ$. 
The lateral translation is therefore treated primarily as a displacement of the focal position.

The two trapping arms were set to orthogonal linear polarizations. 
The degree of polarization of each arm was greater than 99\%, and the polarizing-beam-splitter extinction ratio was greater than $10^3$. 
Under these conditions, residual coherent interference gives less than 0.2\% modulation of the total trapping intensity, much smaller than the force changes produced by focal displacement.

Particle trajectories were obtained from bright-field images after background subtraction. 
The image scale was calibrated by translating a calibration target with a precision stage, giving an effective scale of approximately 170~nm per pixel. 
The particle center in each frame was extracted using an intensity-weighted centroid. 
The resulting time traces were used to calculate position histograms, MSD, PSD, rotation frequency, and projected orbit amplitudes.

Rotation frequencies were extracted from the dominant peak of the PSD of the projected trajectory. 
Unless otherwise noted, the same analysis settings were used for both experimental and simulated data so that the peak frequencies could be compared directly across different alignment conditions. 
For time-resolved frequency analysis, the position trace was divided into consecutive short segments and the dominant PSD peak was extracted from each segment to construct the instantaneous orbital frequency as a function of time.

\section{Optical-force calculation and Langevin simulation}
\label{app:simulation}

The optical force field was calculated using a T-matrix-based optical tweezers toolbox implementing Lorenz--Mie scattering for spherical particles~\cite{Nieminen2007,Lenton2019,Borghese:07}. 
The two trapping beams were modeled as focused Gaussian beams with parameters matched to the measured beam waist and optical power. 
The surrounding medium was taken to be air with refractive index $n_m\simeq1$. 
For comparison with experiment, NaCl-containing aerosols were modeled as spherical particles with effective refractive index $n=1.39$, representing concentrated aqueous particles under the measured humidity conditions.

The calculated optical force was written as
\begin{equation}
\mathbf{F}_{\mathrm{opt}}(\mathbf{r})
=
\frac{n_m P}{c}\mathbf{Q}(\mathbf{r}),
\end{equation}
where $P$ is the single-arm optical power at the objective and $\mathbf{Q}$ is the dimensionless optical force efficiency. 
For each pair of offsets $(\Delta R,\Delta D)$, the force was tabulated on a Cartesian grid and interpolated during trajectory integration.

Most simulations were performed in the plane containing the optical axis and the lateral focal offset. 
This reduced-dimensional treatment captures the dominant coupling between axial scattering imbalance and transverse restoring confinement while allowing efficient scans of alignment space. 
To assess the validity of this approximation, we also performed a limited quasi-three-dimensional calculation by sampling several transverse slices and interpolating $\mathbf{F}(x,y,z)$ during trajectory integration. 
The quasi-three-dimensional model reproduced the same qualitative frequency trends as the reduced model, although near the confinement-to-rotation boundary some trajectories that rotated in the reduced model relaxed to confinement. 
The reduced treatment therefore captures the dominant mechanism and alignment trends, while tending to overestimate the accessible rotation window near regime boundaries.

The particle motion in the reduced $x$--$z$ plane was modeled with an inertial Langevin equation,
\begin{equation}
m\ddot{x}+\gamma \dot{x}=F_x(x,z)+\xi_x(t),\qquad
m\ddot{z}+\gamma \dot{z}=F_z(x,z)+\xi_z(t),
\end{equation}
where $m$ is the particle mass. 
The nominal drag coefficient is
\begin{equation}
\gamma_0=6\pi\eta a,
\end{equation}
for particle radius $a$ in air of viscosity $\eta=1.85\times10^{-5}\,\mathrm{Pa\,s}$. 
In the numerical implementation, a Cunningham slip correction was included so that the effective drag becomes
\begin{equation}
\gamma=\frac{\gamma_0}{C_c}.
\end{equation}
The thermal force satisfies
\begin{equation}
\langle \xi_i(t)\xi_j(t')\rangle
=
2k_BT\gamma\,\delta_{ij}\delta(t-t').
\end{equation}

Trajectories were propagated numerically using substepped integration of the coupled position--velocity dynamics. 
For the inertial model, each substep used an exact underdamped Ornstein--Uhlenbeck update under locally constant force, so that the thermal increments in position and velocity were treated with the correct correlation structure. 
Although the numerical framework supports both overdamped and inertial propagation, the results discussed are based on the inertial Langevin treatment. 
The inclusion of inertia lowers the simulated rotation frequency relative to the overdamped approximation and improves agreement with experiment, although a residual discrepancy remains. 
Likely contributions to this mismatch include uncertainties in particle size, refractive index, beam parameters, and the reduced-dimensional approximation.

The choice of initial conditions can influence the dynamics in weak-confinement and orbital regimes. 
For each parameter set, the initial particle position was therefore randomized uniformly within a small cubic region (\(\pm 2~\mu\mathrm{m}\) in each coordinate) centered on the dual-beam focal center, i.e., the origin of the calculated force map. 
We performed three independent runs for each parameter set using different randomized initial positions and report the majority behavior. 
Accordingly, the simulations are intended to identify representative trajectories and alignment trends, rather than to reproduce ensemble-averaged experimental statistics for every condition.

The time-averaged mean-square displacement was calculated as
\begin{equation}
\mathrm{MSD}(\tau)
=
\left\langle
|\mathbf{r}(t+\tau)-\mathbf{r}(t)|^2
\right\rangle_t ,
\end{equation}
and rotation frequencies were obtained from the dominant peak in the power spectrum of the simulated or measured position trace.

\section{Closed-loop work analysis}
\label{app:close}

For the maps shown in Fig.~\ref{fig:mechanism}, the closed contour $\mathcal{C}$ was chosen as an elliptical loop with size and aspect ratio comparable to the simulated orbit and translated across the analysis plane. The loop work was evaluated numerically as
\begin{equation}
W_{\mathrm{loop}}
\simeq
\sum_n
\mathbf{F}_{\mathrm{opt}}(\mathbf{r}_n)\cdot \Delta \mathbf{r}_n .
\end{equation}

In practice, the closed-loop analysis applies two normalizations. First, the force field is normalized by direction: each force component is divided by the local vector magnitude,
\[
F_{\mathrm{mag}}=\sqrt{F_x^2+F_z^2},
\]
so that only the local force direction is retained and each normalized component lies between $-1$ and $1$. Second, the integrated work is normalized by the loop length, so that the reported quantity is the work per unit path length,
\[
\overline{W}_{\mathrm{loop}}=\frac{W_{\mathrm{loop}}}{L_{\mathcal{C}}},
\]
where $L_{\mathcal{C}}$ is the total contour length. Together, these normalizations allow force fields of different strengths and loop families of different sizes to be compared on the same basis.

To examine the dependence on contour choice, the calculation was repeated for several loop geometries and scaling factors. For each contour family, the reported value corresponds to the maximum normalized closed-loop work obtained after optimizing the loop scale at fixed center.

\section{Confinement benchmark}
\label{sec:best_conf}

\begin{figure}[t]
\centering
\includegraphics[width=\linewidth]{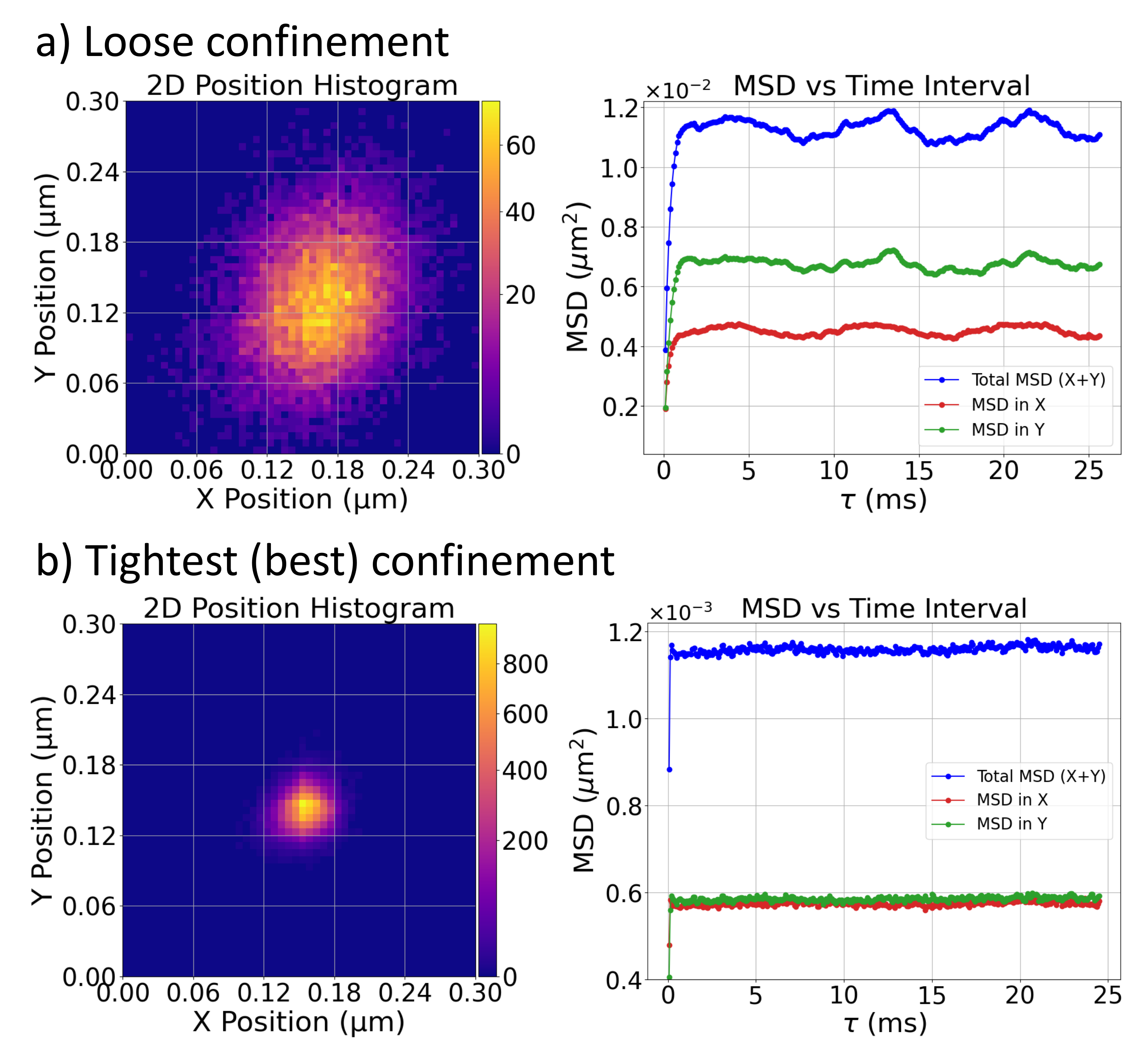}
\caption{
\textbf{Position-stability benchmark.}
(a) Normal confinement condition, showing a broad position distribution and a larger MSD plateau. 
(b) Best confinement after iterative alignment, showing a compact position distribution and a lower MSD plateau.
}
\label{fig:best_conf}
\end{figure}

The attainable confinement was evaluated by comparing a typical alignment condition with the best confinement obtained by minimizing the total MSD. 
In the best-confined state, the position distribution is compact and the MSD reaches a plateau of order \(10^{3}\,\mathrm{nm}^2\). 
The corresponding stiffness was estimated from equipartition,
\begin{equation}
k_x=\frac{2k_B T}{\mathrm{MSD}_x(\infty)}.
\end{equation}
For the measured plateau, this gives \(k_x\sim10^{-5}\,\mathrm{N\,m^{-1}}\), consistent with the stiffness obtained by linearizing the calculated optical force field near the fixed point. 
This agreement indicates that the best-aligned trap operates near the Brownian-noise-limited confinement regime.

\nocite{*}

\bibliography{apssamp}

\end{document}